\documentstyle[emulateapj]{article}
\begin{document}
\submitted{Submitted February 1, 2001; accepted by ApJL}
\title{Diverse Supernova Sources for the $r$-Process and Abundances
in Metal-Poor Stars}
\author{Y.-Z. Qian}
\affil{School of Physics and Astronomy, University of
Minnesota, Minneapolis, MN 55455; qian@physics.umn.edu.}

\begin{abstract}
The dispersion and mean trends of $r$-process abundances 
in metal-poor stars are discussed
based on a model of diverse supernova sources for the $r$-process. 
This model is unique in that its key
parameters are inferred from solar system data independent of stellar 
observations at low metallicities. 
It is shown that this model provides a good
explanation for the observed dispersion and mean trend of Eu abundances
over $-3\lesssim {\rm[Fe/H]}\lesssim -1$. 
It is also shown that this model
provides a means to discuss $r$-abundances in general. 
For example, the Ag abundance 
in any metal-poor star with observed Eu and Fe abundances can be calculated
from the model. This approach is demonstrated with success
for two stars and can be further tested by future Ag data. The dispersion 
and mean trend of Ag abundances in metal-poor stars are also calculated
for comparison with future observations.
\end{abstract}

\section{Introduction}
This paper discusses the dispersion and mean trends of
$r$-process abundances 
($r$-abundances) in ``metal-poor'' stars 
(with $-3\lesssim {\rm [Fe/H]}\lesssim -1$)
based on a model of diverse supernova sources for
the $r$-process inferred from meteoritic data. Wasserburg, Busso, \& 
Gallino (1996) showed that meteoritic data on the $^{129}$I and
$^{182}$Hf inventory in the early solar system require two distinct 
frequencies for replenishment of the $r$-nuclei below
and above mass number $A\sim 130$ in the interstellar medium (ISM). The
frequency required for replenishment of the heavy $r$-nuclei 
($A\gtrsim 130$) is $\sim (10^7\ {\rm yr})^{-1}$ while that for the 
light $r$-nuclei ($A\lesssim 130$) is 
$\sim 10$ times less. These frequencies are consistent 
with the occurrence of Type II supernovae (SNII) 
in an average Galactic ISM
(Wasserburg et al. 1996; see also \S2). The possible realization of 
different $r$-process production in two distinct kinds of
SNII sources, the $\cal{H}$ (high-frequency, mainly producing the
heavy $r$-nuclei) and $\cal{L}$ (low-frequency, mainly producing the
light $r$-nuclei) events, was discussed by Qian, Vogel, \& Wasserburg 
(1998). Based on the solar system data, Qian \& Wasserburg (2000, QW)
calculated the $r$-abundances resulting from a single $\cal{H}$ or 
$\cal{L}$ event. Wasserburg \& Qian (2000, WQ) showed that the results 
of QW were in general agreement with existing observations of 
metal-poor stars.
 
The approach of QW and WQ is further developed here to study the dispersion
in the abundance of a given $r$-element at a given metallicity, with the aim
to gain some new insights into the evolution of $r$-abundances 
relative to Fe and to test SNII as the $r$-process site. 
The element Eu or Ag is chosen as the typical heavy or light 
$r$-element, respectively, for discussion. 
It is shown in \S2 that the model of 
diverse SNII sources for the $r$-process inferred from meteoritic data 
provides a good explanation for the observed dispersion and mean trend of 
Eu abundances in metal-poor stars. It is also shown in \S3 that this model  
provides a means to discuss $r$-abundances in general.
For example, the Ag abundance in any metal-poor star with observed
Eu and Fe abundances can be calculated from the model. This approach is
demonstrated with success for two stars and can be further tested by
future Ag data. The dispersion and mean trend of Ag abundances
in metal-poor stars are also calculated for comparison with future 
observations. Refinements of the model are discussed and conclusions 
given in \S4.

\section{Eu Abundances in Metal-Poor Stars}
Consider the following simple picture for chemical enrichment of 
the Galactic ISM
by SNII (e.g., Qian 2000). The ejecta of any individual SNII is mixed with
a fixed amount of ISM corresponding to the total mass swept up by an SNII
remnant, $M_{\rm mix}\approx 3\times 10^4\,M_\odot$ (e.g., Thornton et al. 
1998). The SNII rate per unit mass of gas $f_{\rm G}^{\rm SN}/M_{\rm gas}$
is assumed to be constant over Galactic history. Consequently, an average 
ISM in the Galaxy is enriched in the nucleosynthetic products from SNII at 
a frequency
\begin{eqnarray}
f_{\rm mix}^{\rm SN}&=&M_{\rm mix}{f_{\rm G}^{\rm SN}\over M_{\rm gas}}
=(10^7\ {\rm yr})^{-1}\nonumber\\
&\times&\left({M_{\rm mix}\over 3\times 10^4\,M_\odot}\right)
\left[{f_{\rm G}^{\rm SN}\over (30\ {\rm yr})^{-1}}\right]
\left({10^{10}\,M_\odot\over M_{\rm gas}}\right),
\label{fmix}
\end{eqnarray}
where $f_{\rm G}^{\rm SN}/M_{\rm gas}$
is estimated using quantities for the present Galaxy. The frequency of
$\cal{H}$ events inferred from the meteoritic data is
$f_{\cal{H}}\sim f_{\rm mix}^{\rm SN}$ and that of $\cal{L}$ events
is $f_{\cal{L}}\sim 0.1f_{\rm mix}^{\rm SN}$. For definiteness, 
$f_{\cal{H}}=(10^7\ {\rm yr})^{-1}$ and 
$f_{\cal{L}}=(10^8\ {\rm yr})^{-1}$ are used below. It is further
assumed that all SNII are either $\cal{H}$ or $\cal{L}$ events, with the
average fraction of each kind being 
$q=f_{\cal{H}}/(f_{\cal{H}}+f_{\cal{L}})=10/11$ for $\cal{H}$ events 
or $1-q=1/11$ for $\cal{L}$ events. The nucleosynthetic yields in an
$\cal{H}$ or $\cal{L}$ event are taken to be constant.

Over the Galactic history of $\approx 10^{10}$~yr before solar system 
formation, $\approx 10^3$ $\cal{H}$ and $\approx 10^2$ $\cal{L}$ 
events contributed to the solar abundances. As the $\cal{H}$ events
are mainly responsible for the heavy $r$-elements such as Eu, the Eu
abundance resulting from a single $\cal{H}$ event must be
$({\rm Eu/H})_{\cal{H}}\approx 10^{-3}({\rm Eu/H})_{\odot,r}$, where
Eu/H is the number ratio of Eu to hydrogen and 
$({\rm Eu/H})_{\odot,r}\approx 3.29\times 10^{-12}$ is the  
solar $r$-process value (Arlandini et al. 1999). In the spectroscopic 
notation, $\log\epsilon_{\cal{H}}({\rm Eu})
\equiv\log({\rm Eu/H})_{\cal{H}}+12\approx -2.48$. By ignoring possible Eu
contribution from the $\cal{L}$ events, 
the Eu abundance in a star can be considered as resulting from 
a number $n_{\cal{H}}$ of $\cal{H}$ events:
\begin{equation}
\label{eu}
\log\epsilon({\rm Eu})\approx \log\epsilon_{\cal{H}}({\rm Eu})
+\log n_{\cal{H}}.
\end{equation}
For example, HD 122563, 
HD 115444, and CS 22892-052 with [Fe/H]~$= -2.74$, $-2.99$, and $-3.1$ 
have $\log\epsilon({\rm Eu})=-2.59$, $-1.63$, and $-0.93$, respectively
(Westin et al. 2000; Sneden et al. 2000), corresponding to 
$n_{\cal{H}}\approx 1$, 7, and 35.

The three stars mentioned above all have [Fe/H]~$\approx -3$ but very
different $\log\epsilon({\rm Eu})$ values. In fact, this large dispersion
in $\log\epsilon({\rm Eu})$ at [Fe/H]~$\approx -3$ has been demonstrated by
more extensive observations (McWilliam et al. 1995; Burris et al. 2000). 
Application of equation (\ref{eu}) to the Eu 
data at [Fe/H]~$\approx -3$ led WQ to conclude that very little Fe can be 
produced by the $\cal{H}$ events. It was shown that $\approx 1/3$ of 
the solar Fe inventory was provided by SNII (e.g., Timmes, Woosley, \& 
Weaver 1995; Qian \& Wasserburg 2001a). This Fe must be assigned to the 
$\approx 10^2$ $\cal{L}$ events that contributed to the solar abundances. 
Consequently, the Fe abundance resulting from a single $\cal{L}$ event must 
be $({\rm Fe/H})_{\cal{L}}\approx ({\rm Fe/H})_{\odot}/300$, corresponding 
to [Fe/H]$_{\cal{L}}\equiv\log({\rm Fe/H})_{\cal{L}}-
\log({\rm Fe/H})_{\odot}\approx -2.48$. Furthermore, WQ concluded that
the Fe inventory at [Fe/H]~$\approx -3$ must be provided by the first
very massive stars (with masses of $\gtrsim 100 M_\odot$) before any
SNII could have occurred. These stars may have produced other elements
along with Fe, thus providing an ``initial'' inventory of metals to the ISM.
However, data on Ba (a heavy $r$-element)
at $-4\lesssim [{\rm Fe/H}]\lesssim -3$ 
(McWilliams et al. 1995; McWilliam 1998) show that the initial inventory
of heavy $r$-elements is negligible compared with the contribution from a
single $\cal{H}$ event (WQ). Therefore, 
the initial Eu inventory can be ignored in
equation (\ref{eu}) for $\log\epsilon({\rm Eu})$. The initial inventory of 
primary concern here is that of Fe, hereafter denoted as 
[Fe/H]$_0=-3$. As Type Ia supernovae (SNIa) contributed Fe only at 
times corresponding to [Fe/H]~$\gtrsim -1$ (e.g., Timmes et al. 1995; 
Qian \& Wasserburg 2001a), the Fe abundance in a
metal-poor star with $-3\lesssim [{\rm Fe/H}]\lesssim -1$ can be considered
as resulting from a number $n_{\cal{L}}$ of $\cal{L}$ events:
\begin{equation}
\label{fe}
[{\rm Fe/H}]\approx\log\left(10^{[{\rm Fe/H}]_0}+n_{\cal{L}}\times
10^{[{\rm Fe/H}]_{\cal{L}}}\right).
\end{equation}
The range $-3\lesssim [{\rm Fe/H}]\lesssim -1$
of interest here corresponds to $n_{\cal{L}}\approx 0$--30.

Based on the above discussion, the Eu and Fe abundances in a metal-poor
star can be considered as resulting from a mixture of $\cal{H}$ and
$\cal{L}$ events. Given the average fractions of $\cal{H}$ and
$\cal{L}$ events among all SNII, the distribution of 
$\log\epsilon({\rm Eu})$ expected for metal-poor stars with a given
[Fe/H] can be calculated. Consider the occurrence of SNII in an ISM 
with the average 
fraction of $\cal{H}$ events being $q$. The probability to have a number
$n_{\cal{H}}$ of $\cal{H}$ events mixed with a number $n_{\cal{L}}$ of
$\cal{L}$ events is the same as that for the 
$(n_{\cal{H}}+n_{\cal{L}}+1)$th SNII to be the $(n_{\cal{L}}+1)$th
$\cal{L}$ event. The latter is given by the so-called negative binomial
distribution:
\begin{equation}
\label{bi}
P(n_{\cal{H}},n_{\cal{L}})={(n_{\cal{H}}+n_{\cal{L}})!\over 
n_{\cal{H}}!n_{\cal{L}}!}(1-q)^{n_{\cal{L}}+1}q^{n_{\cal{H}}}.
\end{equation}
In particular, $P(n_{\cal{H}},n_{\cal{L}}=0)=
(1-q)q^{n_{\cal{H}}}$. Thus the mixtures of $(n_{\cal{H}},n_{\cal{L}})
\approx(1,0)$, (7,0), and (35,0) corresponding to the Eu and Fe 
abundances in HD 122563, HD 115444, and CS 22892-052 would occur with
probabilities of $\approx 8\%$, 5\%, and 0.3\% at [Fe/H]~$\approx -3$
for $q=10/11$. In general, among all the mixtures with a given 
$n_{\cal{L}}$ the fraction having $n_{\cal{H}}\leq n_{\cal{H}}^x$ is
\begin{equation}
\label{x}
x=\sum_{n_{\cal{H}}=0}^{n_{\cal{H}}^x}P(n_{\cal{H}},n_{\cal{L}}).
\end{equation}
Consequently, a fraction $x$ of all the stars with [Fe/H] given by
equation (\ref{fe}) for the same $n_{\cal{L}}$ as in equation (\ref{x})
should have $\log\epsilon({\rm Eu})\leq\log\epsilon^x({\rm Eu})
\approx\log\epsilon_{\cal{H}}({\rm Eu})+\log n_{\cal{H}}^x$. 
An ``average'' value of $\log\epsilon({\rm Eu})$
at a given [Fe/H] can be defined as
$\langle\log\epsilon({\rm Eu})\rangle\approx
\log\epsilon_{\cal{H}}({\rm Eu})+\log\langle n_{\cal{H}}\rangle$, where
\begin{equation}
\langle n_{\cal{H}}\rangle={q\over 1-q}(n_{\cal{L}}+1)
\end{equation}
is the mean value of $n_{\cal{H}}$ for a given $n_{\cal{L}}$ according to
the distribution in equation (\ref{bi}).

By taking $\log\epsilon_{\cal{H}}({\rm Eu})=-2.48$, [Fe/H]$_0=-3$,
[Fe/H]$_{\cal{L}}=-2.48$, and $q=10/11$, 
the contours of $\log\epsilon^x({\rm Eu})$ are shown in Figure 1 as functions 
of [Fe/H] for $x\approx 1\%$, 10\%, 90\%, and 99\%. These can be 
interpreted as the 99\% (solid lines, $x\approx 1\%$ and 99\%) or 90\% 
(dashed lines, $x\approx 10\%$ and 90\%) ``confidence contours'' for Eu 
abundance at a given [Fe/H]. 
If the model presented here is correct, then
it should be very unlikely to find stars with values of 
$\log\epsilon({\rm Eu})$ and [Fe/H] outside the 99\% confidence contours.
The dot-dashed line in Figure 1 shows 
$\langle\log\epsilon({\rm Eu})\rangle$ as a function of [Fe/H] and can be 
taken as the ``mean'' trend for evolution of Eu abundance relative to Fe.
The Eu data of McWilliam et al. (1995), Westin et al. (2000), 
Sneden et al. (2000), and Burris et al. (2000) are also shown in Figure 1.
The data at $-3\lesssim [{\rm Fe/H}]\lesssim -1.3$ appear to be described 
quite well by the confidence contours and mean trend calculated from
the model, especially when typical observational errors of 
$\approx \pm 0.1$--0.2 for $\log\epsilon({\rm Eu})$ are taken into 
account. As the model only considers Fe addition from the $\cal{L}$ events
to the initial Fe inventory, it tends to overestimate $n_{\cal{L}}$ [hence
$n_{\cal{H}}^x$ and $\log\epsilon^x({\rm Eu})$] near 
the onset of SNIa Fe contribution at [Fe/H]~$\sim -1$. This explains the 
few data points that are significantly below the contour for 
$x\approx1\%$ at [Fe/H]~$\gtrsim -1.3$. These data points are expected to 
be inside the 99\% confidence contours when SNIa Fe contribution is 
allowed for. The vertical segments of the contours for
$x\approx 1\%$ and 10\% are caused by $n_{\cal{H}}^x=0$ 
[$\log\epsilon^x({\rm Eu})=-\infty$] for $n_{\cal{L}}=1$ and 0, 
respectively. These segments will be modified by a refined model that takes
into account a small initial Eu inventory associated with 
[Fe/H]$_0=-3$ and possible Eu contribution
from the $\cal{L}$ events (see \S4). 

\section{Ag Abundances in Metal-Poor Stars}
As discussed in \S2, the model of diverse SNII sources for the $r$-process
based on the characteristics of $\cal{H}$ and $\cal{L}$ events inferred
from meteoritic data provides a good explanation for the observed
dispersion and mean trend of Eu abundances in metal-poor stars. 
There may be other 
possible models that can achieve the same by separating Eu and Fe
production in a similar way. However, the model presented here is unique
in that its key parameters are inferred from solar system data independent of
stellar observations at low metallicities. Furthermore, this model 
provides a means to discuss $r$-abundances in general. For example, 
the abundance of Ag (a light $r$-element) in a metal-poor star 
can be considered as 
resulting from a mixture of $\cal{H}$ and $\cal{L}$ events in addition to
a possible initial Ag inventory associated with [Fe/H]$_0=-3$.
The numbers $n_{\cal{H}}$ and $n_{\cal{L}}$ for the mixture can be 
determined from the observed Eu and Fe abundances as discussed in \S2.
Given the initial Ag inventory $\log\epsilon_0({\rm Ag})$, the Ag abundance
resulting from a single $\cal{H}$ event $\log\epsilon_{\cal{H}}({\rm Ag})$,
and that from a single $\cal{L}$ event $\log\epsilon_{\cal{L}}({\rm Ag})$,
the Ag abundance in a metal-poor star can be calculated as
\begin{eqnarray}
\log\epsilon({\rm Ag})&\approx&
\log\left[10^{\log\epsilon_0({\rm Ag})}+
n_{\cal{H}}\times 10^{\log\epsilon_{\cal{H}}({\rm Ag})}\right.\nonumber\\ 
&+&\left.n_{\cal{L}}\times 10^{\log\epsilon_{\cal{L}}({\rm Ag})}\right].
\label{ag}
\end{eqnarray}

Stars with [Fe/H]~$\approx -3$ correspond to mixtures with 
$n_{\cal{L}}\approx 0$ (see eq. [\ref{fe}]).
If Ag abundances are known for two such stars with different observed
Eu abundances (hence different $n_{\cal{H}}$, e.g., HD 115444 and 
CS 22892-052), then the parameters $\log\epsilon_0({\rm Ag})$ and
$\log\epsilon_{\cal{H}}({\rm Ag})$ can be obtained from two equations 
similar to equation (\ref{ag}). However, so far the
Ag abundance in only one star with [Fe/H]~$\approx -3$ has been
published [$\log\epsilon({\rm Ag})=-0.80$ for CS 22892-052, 
Sneden et al. 2000]. For the discussion to proceed, 
it is assumed that the initial Ag inventory
can be ignored [$\log\epsilon_0({\rm Ag})=-\infty$]. With
$(n_{\cal{H}},n_{\cal{L}})\approx (35,0)$ for CS 22892-052 (see \S2), 
$\log\epsilon_{\cal{H}}({\rm Ag})\approx -2.34$ is obtained. As the solar 
$r$-abundance of Ag resulted from $\approx 10^3$ $\cal{H}$
and $\approx 10^2$ $\cal{L}$ events (see \S2), an equation 
similar to equation (\ref{ag}) also applies to the sun.
The total solar photospheric
abundance $\log\epsilon_\odot({\rm Ag})=1.08$ (Crawford et al. 1998)
and the solar $s$-process abundance 
(Ag/H)$_{\odot,s}=3.45\times 10^{-12}$ (Arlandini et al. 1999) correspond 
to $\log\epsilon_{\odot,r}({\rm Ag})\approx 0.93$, which gives
$\log\epsilon_{\cal{L}}({\rm Ag})\approx -1.40$ (the Ag yield in an
$\cal{L}$ event is $\approx 9$ 
times higher than that in an $\cal{H}$ event,
as expected from the characteristics of these events). With the parameters
$\log\epsilon_0({\rm Ag})$, $\log\epsilon_{\cal{H}}({\rm Ag})$, and
$\log\epsilon_{\cal{L}}({\rm Ag})$ chosen above, the Ag abundance 
in any metal-poor star with observed Eu
and Fe abundances can be calculated by using equations (\ref{eu}),
(\ref{fe}), and (\ref{ag}).

Crawford et al. (1998)
reported Ag abundances or upper limits for 7 stars with 
[Fe/H]~$\approx -2.46$ to $-1.32$, two of which also have observed Eu
abundances from Burris et al. (2000). For these two stars the Ag abundances 
calculated from the above model are in good agreement with
the observed values (see Table 1; the observational error for 
$\log\epsilon({\rm Ag})$ is $\approx\pm 0.2$). The model can be further tested
by future Ag data. As part of the test,
stars that are predicted to have high Ag abundances can be selected first
for observation/analysis. For stars with observed Ag abundances
but no Eu data, Figure 2 shows the dispersion and mean trend
of Ag abundances over $-3\lesssim{\rm [Fe/H]}\lesssim -1$ 
calculated as in the case of Eu (see \S2) 
for $\log\epsilon_0({\rm Ag})=-\infty$,
$\log\epsilon_{\cal{H}}({\rm Ag})=-2.34$,
$\log\epsilon_{\cal{L}}({\rm Ag})=-1.40$, and $q=10/11$. 
The Ag data of Crawford et al. (1998) and Sneden et al. (2000) are also
shown in Figure 2. While the available data are fully consistent with the
model, extensive Ag data for metal-poor 
stars are needed for a detailed comparison.
The vertical segments of the confidence contours in Figure 2
are caused by the assumption of $\log\epsilon_0({\rm Ag})=-\infty$ and will
be modified when the actual value of 
$\log\epsilon_0({\rm Ag})$ is known (see \S4).

\section{Discussion and Conclusions}
The dispersion and mean trends of $r$-abundances 
in metal-poor stars have been discussed
based on a model of diverse SNII sources for the $r$-process. 
This model is unique in that its key
parameters are inferred from solar system data independent of stellar 
observations at low metallicities. It has been shown that this model provides 
a good explanation for the observed dispersion and mean trend of Eu 
abundances over $-3\lesssim {\rm [Fe/H]}\lesssim -1$. The calculated
confidence contours and mean trend for Eu in Figure 1 and those for Ag
in Figure 2 represent a good test to distinguish SNII from neutron
star mergers as the $r$-process site. Extremely high Eu and Ag abundances
compared with the confidence contours in Figures 1 and 2 would be 
expected for metal-poor stars if neutron star mergers were the $r$-process
site (Qian 2000). If the results presented here
are borne out by more extensive Eu and Ag data,
then it is quite secure to conclude that SNII are the major sources
for the $r$-process. 
Furthermore, the model presented here
provides a means to discuss $r$-abundances in general. 
For example, the Ag abundance in any metal-poor star with observed Eu and Fe 
abundances can be calculated from the model. 
This approach has been 
demonstrated with success for two stars (Table 1) 
and can be further tested by
future Ag data. The full approach for a large number of
$r$-elements will be discussed in a separate paper 
(Qian \& Wasserburg 2001b).

As emphasized above, the parameters for the model presented here are
inferred from solar system data. In particular, the fraction $q$ of
$\cal{H}$ events among all SNII is estimated from the frequencies of
$\cal{H}$ and $\cal{L}$ events suggested by the meteoritic data on $^{129}$I
and $^{182}$Hf. While the specific choice of $q=10/11$ appears to describe
the existing observations quite well [this is confirmed by the independent
work of Fields, Truran, \& Cowan (2001)], the actual value of $q$ may be
obtained from e.g., an extensive set of Eu data at [Fe/H]~$\approx -3$.
Stars with [Fe/H]~$\approx -3$ 
contain no contributions from the $\cal{L}$ events.
The probability for such stars to have received contributions from a number
$n_{\cal{H}}$ of $\cal{H}$ events is simply proportional to $q^{n_{\cal{H}}}$
(\S2). Abundances of Eu in these stars can be converted into a histogram on
the frequency of occurrences for different $n_{\cal{H}}$. 
Then the value of $q$
can be obtained by an exponential fit to the histogram. It can be seen from
Figure 1 that the existing Eu data at [Fe/H]~$\approx -3$ are rather 
inadequate for this purpose. Future observations in this region of [Fe/H]
are highly desirable.

The initial Eu inventory associated with [Fe/H]$_0=-3$ and possible Eu 
contribution from the $\cal{L}$ events have been ignored.
This gives rises to the vertical segments of the confidence contours
in Figure 1. The actual initial Eu inventory may be estimated from the
Eu abundances at [Fe/H]~$<-3$ corresponding to the epoch for production
by the first very massive stars. The actual Eu production in an $\cal{L}$ 
event is more complicated to determine. One possibility is to detect
overabundances of $r$-elements on the surface of a binary companion to
a neutron star (Qian 2000). If the SNII producing the
neutron star was an $\cal{L}$ event, then large overabundances 
of light $r$-elements such as Ag should be observed.
Any associated overabundance of Eu, if detected, can be used to estimate
the Eu production in an $\cal{L}$ event.

The initial Ag inventory have also been ignored by taking
$\log\epsilon_0({\rm Ag})=-\infty$. The actual value of  
$\log\epsilon_0({\rm Ag})$ can be
determined when the Ag abundance in another star with 
[Fe/H]~$\approx -3$ (e.g., HD 115444) is known. The upper limit
on the Ag abundance in HD 140283 with [Fe/H]~$=-2.46$ (Crawford et al. 1998) 
indicates that $\log\epsilon_0({\rm Ag})\lesssim -1.15$ (see Figure 2). 
In any case, extensive Ag data are needed to test and refine the 
model presented here.

\acknowledgments
I thank Jerry Wasserburg for many discussions on Galactic chemical evolution, 
Friedel Thielemann for calling my attention to the work of Burris et al. 
(2000), and Evan Skillman, Keith Olive, and an anonymous referee for 
suggestions that improve the paper. I also acknowledge Chris Sneden, 
John Cowan, and their collaborators for providing continuous stimulus by 
carrying out extensive observations. This work was supported in part by the 
Department of Energy under grants DE-FG02-87ER40328 and DE-FG02-00ER41149.

\figcaption{The dispersion and mean trend of Eu abundances in metal-poor 
stars calculated from the model are compared with data (open squares: 
McWilliam et al. 1995; asterisks: Westin et al. 2000; filled diamond:
Sneden et al. 2000; filled circles: Burris et al. 2000). The solid and
dashed lines correspond to the calculated 99\% and 90\% confidence 
contours for Eu abundance at a given [Fe/H]. The dot-dashed line is the 
calculated mean trend for evolution of Eu abundance relative to Fe.}

\figcaption{Same as Figure 1, but for Ag. Open triangles (upper limits)
and filled squares are new symbols indicating data from Crawford et al.
(1998).}

\begin{deluxetable}{crrrrrr}
\footnotesize
\tablecaption{Comparison of calculated Ag abundances with data}
\tablewidth{0pt}
\tablehead{
\colhead{Star}&\colhead{[Fe/H]}&\colhead{$n_{\cal{L}}$}&
\colhead{$\log\epsilon({\rm Eu})$}&\colhead{$n_{\cal{H}}$}& 
\colhead{$\log\epsilon_{\rm obs}({\rm Ag})$}&
\colhead{$\log\epsilon_{\rm cal}({\rm Ag})$}
}
\startdata
HD 2665&$-1.95$&3&$-1.23$&18&$-0.88$&$-0.70$\\
HD 6755&$-1.60$&7&$-0.56$&83&$-0.16$&$-0.18$\\
\enddata
\tablecomments{Data on Fe and Ag [$\log\epsilon_{\rm obs}({\rm Ag})$]
are taken from Crawford et al. (1998), and data on Eu from Burris et al. 
(2000). The $\log\epsilon_{\rm cal}({\rm Ag})$ values are
calculated from the model (see text).}
\end{deluxetable}

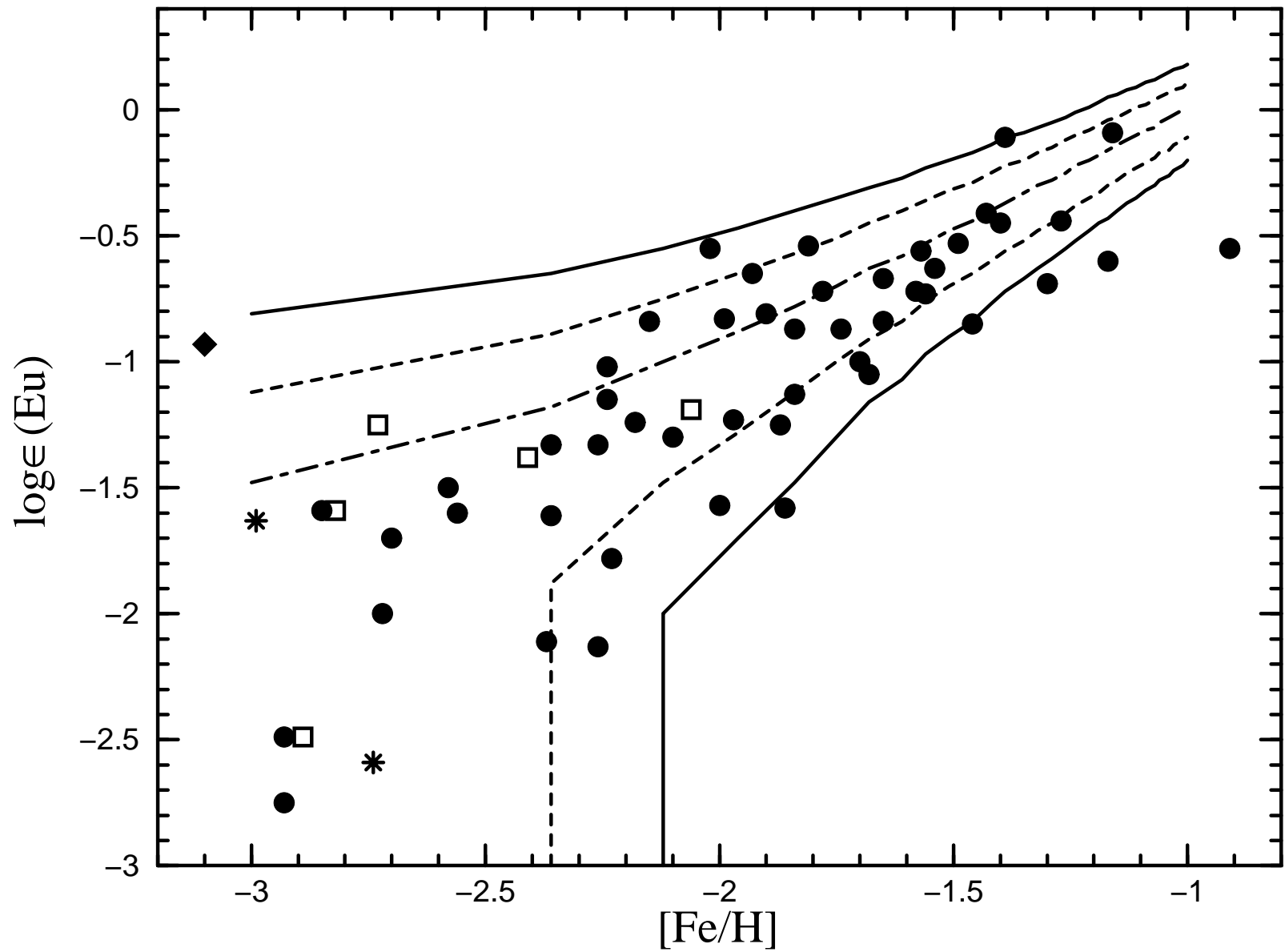
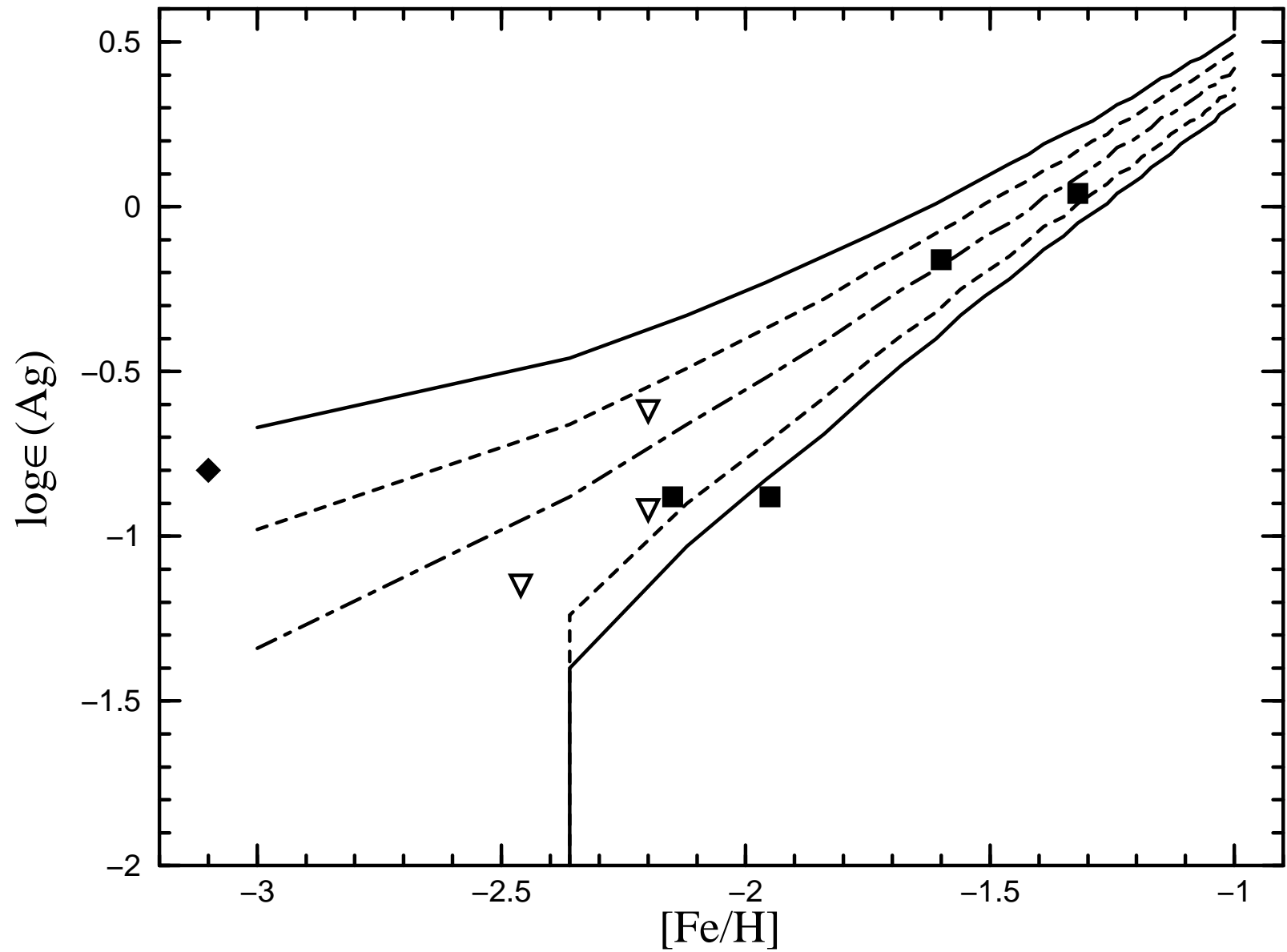
\end{document}